\documentclass[aps,prb,twocolumn,showpacs,twoside,amsmath, amsfonts,floatfix]{revtex4-1}
\usepackage{psfrag}
\usepackage{graphicx}
\usepackage{epsfig}
\usepackage{amssymb}
\usepackage{color}

\newcommand{\myred}{}

\begin{document}
\title{
L\'evy flights and multifractality in quantum critical diffusion \protect\\
and in classical random walks on fractals
%
%
}
\author {V.~E.~Kravtsov$^{1,2}$, O.~M.~Yevtushenko$^{3}$, P.~Snajberk$^{3}$ and E.~Cuevas$^{4}$ }
\affiliation{
   $^{1}$The Abdus Salam International Centre for
             Theoretical Physics, P.O.B. 586, 34100 Trieste, Italy \\
   $^{2}$Landau Institute for Theoretical Physics, 2 Kosygina
             st., 117940 Moscow, Russia \\
   $^3$Arnold Sommerfeld Center for Theoretical Physics and Center for Nano-Science, Ludwig-Maximilians-University, 80333 Munich, Germany \\
   $^4$Departamento de F{\'i}sica, Universidad de Murcia, E30071 Murcia, Spain
}

\date{today}
\begin{abstract}
We employed the method of virial expansion in order to compute the
retarded density correlation function (generalized diffusion
propagator) in the critical random matrix ensemble in the limit of
strong multifractality. We found that the long-range nature of the
Hamiltonian is a common root of both multifractality and L\'evy
flights which show up in the power-law intermediate- and
long-distance behavior, respectively, of the  density correlation
function. We review certain models of classical random walks on
fractals and show the similarity of the density correlation function
in them to that for the quantum problem described by the random
critical long-range Hamiltonians.

\end{abstract}

\renewcommand{\(}{\left(}
\renewcommand{\)}{\right)}
\newcommand{\la}{\left\langle}
\newcommand{\ra}{\right\rangle}

\setlength{\parindent}{0em}

\date{\today}

\maketitle
\section{Introduction}
Multifractality of critical wave functions \cite{Weg} has recently
emerged as a subject of not only fundamental theoretical interest
\cite{Mirlin-rev} but also of experimental activities at a frontier
of technological achievements \cite{Huse, Faez, Delande} in surface
imaging, sound propagation and driven cold atom systems. It could be
also an important issue in describing the superconductor (superfluid)
to insulator transitions in condensed matter and systems of cold
atoms \cite{FIK}, Kondo effect in strongly disordered metals
\cite{KetIm} and other interaction and non-linear phenomena in
systems with strong disorder. Multifractal correlations is the
property of wave functions near the Anderson localization transition
in 3D  systems  
with short-range Hamiltonians as well as in
 systems  with long-range $|H_{r,r'}|\sim |r-r'|^{-d}$ hopping
\cite{MFSeil,KrMut,LevAl,EfAl}; {\myred $ \, d \, $ is the space dimension}. 
In the latter systems critical correlations of wave functions may be 
realized in a broad range of parameters.

Usually multifractality is studied by considering the moments of the
local density of states at a given energy $\varepsilon$ or the
correlation functions of such moments at different energies.
However, it may show up also in the density correlation function:
\begin{equation}
\label{D-r} {\cal \tilde{D}}_{r,r'}(\omega)=\left\langle
\sum_{n,m}\frac{\psi_{n}(r')\psi_{n}^{*}(r)\psi_{m}(r)\psi_{m}^{*}(r')}
{(\varepsilon-\varepsilon_{n}+i0)\,(\varepsilon-\varepsilon_{m}-\omega
-i0)}\right\rangle.
\end{equation}
In Eq. (\ref{D-r}) we denote by $\psi_{n}(r)$ the exact $n$-th wave
function at a site $r$ which corresponds to the energy
$\varepsilon_{n}$ and $\varepsilon$ is the fixed energy (which we
set zero throughout this paper). The symbol $\langle...\rangle$
stands for the ensemble averaging.
\begin{figure}[ht]
 \includegraphics[width=8.5cm, 
                             angle=0]{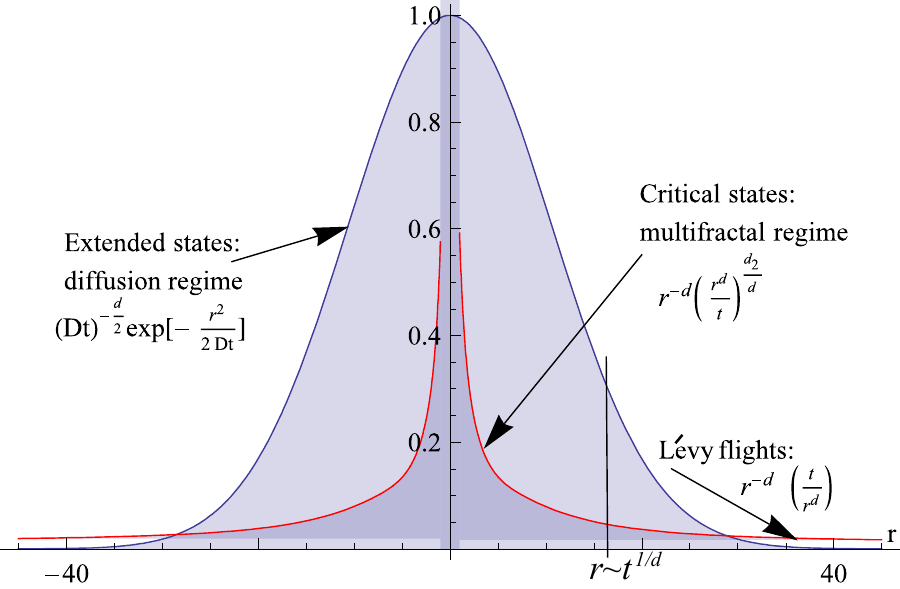}
\caption{
(color online) Generalized diffusion propagator (density
correlation function) for critical eigenstates of long-range
Hamiltonians Eq. (\ref{long-range}) with strong multifractality
$d_{2}\ll d$. The multifractality is encoded in the
intermediate-distance 
%
%
($1\ll r/\ell_{0} \ll (E_{0}\,t)^{\frac{1}{d}} $) 
dependence $r^{-d+d_{2}}$ while the L\'evy flights exhibit
themselves in the long-distance 
%
%
($r/\ell_{0}\gg (E_{0}\,t)^{\frac{1}{d}}$) 
tails $r^{-2d}$;
%
%
$\ell_{0}$, $E_{0}$ are the lower length scale and the upper energy scale 
of multifractality \cite{KravCue}, respectively. The overall functional
form of the density correlation function obeys the scaling law
$r^{-d}\,S(r^{d}/t)$.
} \label{Levy}
\end{figure}
The Fourier transform $\hat{{\cal D}}_{r,r'}(t)$ of this correlation
function (``generalized diffusion propagator'') describes spreading of
the wave packet in space and time and is a direct generalization of
the pure classical object: the probability density to find a random
walker at a point $r$ at a time $t$ after the beginning of random
walk from the origin{\myred, $ \, r'=0$}. The possibility of both classical and quantum
formulation of the density correlation function makes this object
convenient for quantum-to-classical comparison.

The descender of the generalized diffusion propagator is the {\it
survival probability} or the {\it return probability} $P(t)$
\begin{equation}
\label{P-t} 2\pi\rho_{0}\,P(t)= \hat{{\cal D}}_{r,r}(t),
\end{equation}
where $\rho_{0}$ is the averaged DoS.

This is the probability for a wave packet to survive for a time $t$
at a point where it was created at $t=0$ or the probability for a
random walker to find himself in the origin at a time $t$. For
extended states $P(t)$ decreases with time, albeit differently for
the truly extended states and the critical multifractal states
\cite{Chalker, Chalker-proof}:
\begin{equation}
\label{P-multi} P(t)\propto \left\{\begin{matrix} t^{-\frac{d}{2}},&
{\rm extended\;\;\; states}\cr t^{-\frac{d_{2}}{d}},& {\rm
critical\;\;\; states}\end{matrix}\right.
\end{equation}
In Eq.(\ref{P-multi}), $d_{2}<d$ is the non-trivial critical
exponent, the correlation fractal dimension, that determines inverse
participation ratio in a critical system of the size $L$:
\begin{equation}
\label{IPR} {\cal I}_{2}=\left\langle \sum_{r}|\psi(r)|^{4} \right\rangle
\propto L^{-d_{2}}.
\end{equation}
A general form of the density correlation function, $ \, {\cal
D}(\kappa,\omega) $, was suggested by Chalker  in
the momentum-frequency representation \cite{Chalker, Ch-D}. 
In the region of critical states it can be written as follows:
\begin{equation}
\label{Ch-cr} 
{\cal D}(\kappa,\omega)=\frac{2\pi\rho_{0}}{(-i\omega+0)\,[f(|\kappa|\,L_{-i\omega})+1]},
\end{equation}
where $\rho_{0}=N^{-1}\sum_{n=0}^{N}\langle \delta(\varepsilon-\varepsilon_{n})\rangle|_{\varepsilon=0}$
is the averaged density of states at
$\varepsilon=0$.

The key element of this prediction is that at criticality (in the
classical Dyson symmetry classes) there is only one scale
$1/L_{\omega}\propto \omega^{\frac{1}{d}}$ which separates small-
and large-momentum behavior, both of them being different limits of
one and the same analytic function $f(x)$. The combination
$L_{-i\omega}\propto (-i\omega+0)^{-d}$ in Eq.(\ref{Ch-cr}) is
dictated by the {\it retarded} character of the correlation function
which should be a regular function in the upper half-plane of
complex $\omega$.

Note that the original idea of Chalker was that of reconciling the one-parameter 
scaling of Gang-of-Four \cite{GoF} with the multifractal behavior. The one-parameter 
scaling is encoded in Eq.(\ref{Ch-cr}): the counterpart of $  {\cal D}(\kappa,\omega) $ 
in the coordinate-time domain obeys the scaling law
\begin{equation}
\label{scaling}
    r^{d}\,\hat{{\cal D}}_{r,0}(t)= S(r^{d}/t)\,,
%
%
\end{equation}
where $S(x)$ is the scaling function.

According to {\it Chalker's conjecture}, the multifractality
exhibits itself in the asymptotic form of the function $f(x)$ at
$|x|\gg 1$:
\begin{equation}
\label{Ch-conj} f(x)=c\, x^{d_{2}}, \;\;\;\;\;|x|\gg 1.
\end{equation}
Here $d_{2}$ is the same correlation fractal dimension defined by
Eq. (\ref{IPR}) which appears also in Eq. (\ref{P-multi}) as well as
in the local density of states correlation function Ref.\cite{Chalker-proof}. 
It determines the {\it multifractal regime} of the density correlation function (see Fig. 1) where
\begin{equation}
\label{MF-regime} \hat{{\cal D}}_{r,0}(t)\propto r^{-d}\,\left(
\frac{r^{d}}{t}\right)^{\frac{d_{2}}{d}},\;\;\;\;\;(t>r^{d})\,.
\end{equation}
At small $|x|\ll 1$, the function $f(x)$ is small as the probability
conservation
\begin{equation}
\label{sum-rule} \sum_{r'}\hat{{\cal
D}}_{r,r'}(t)=2\pi\rho_{0}\,\theta(t)
\end{equation}
requires
\begin{equation}
\label{k-0} f(x=0)=0\,.
\end{equation}
Assuming that $x=0$ is the regular point of the analytic function
$f(x)$, one can expand
\begin{equation}
\label{non-anal}
f(x)=c_{1}\,x+c_{2}\,x^{2}+c_{3}\,x^{3}+...,\;\;\;\;(x=|\kappa|\,L_{-i\omega}).
\end{equation}
Note, however, that the regular expansion in $x$, Eq.(\ref{non-anal}),
does not mean analyticity in the momentum domain. For instance the
linear in $x$ term contains {\it non-analytic} term $|\kappa|$.
Non-analytic in $\kappa$ terms arise from any {\it odd} power of
$x$. These terms imply power-law tails of the density correlation
function in the coordinate domain. Namely, if the leading odd term
in Eq. (\ref{non-anal}) is $x^{2m-1}$, the density correlation
function acquires a power-law tail $\propto r^{-(2m-1+d)}$ at large
$r$. By analogy with classical random walks, we will refer to such
{\myred tails} as the {\it L\'evy flights}.

{\myred
The simplest class of critical systems, where such non-analytic
terms are present, includes the systems with long-range
resonance hopping where the hopping terms of the Hamiltonian 
decrease with distance as:
}
%
%
\begin{equation}
\label{long-range} |\hat{H}_{r,r'}|\propto |r-r'|^{-d}.
\end{equation}
However, the situation depends a lot on the dimensionality $d$. As
we will show below, the L\'evy flights in such systems are
characterized by a power-law tail:
\begin{equation}
\label{PL-tail} 
   \hat{{\cal D}}_{r,0}(t)\propto r^{-d}\,\left( \frac{t}{r^{d}}\right),\;\;\;\;\;(t<r^{d})\,.
\end{equation}
For $d=3$, the second moment $\langle r^{2}\rangle$ does exist so
that we have in Eq. (\ref{non-anal}) that $c_{1}=0$ and $c_{2}$ is finite.
The non-analytic in ${\kappa}$ terms appear starting from $x^{3}$,
which, however, is sub-leading at small $x$. The role of
non-analyticity is greatly enhanced at $d=1$ where $c_{1}\neq 0$.
Here the leading term in Eq. (\ref{non-anal}) is non-analytic in
$\kappa$. The case $d=2$ is special \cite{EfAl}. In this case even
criticality is not guaranteed by Eq. (\ref{long-range}) and the
answer may depend on the particular symmetry class and details of
the Hamiltonian.

In this paper, we concentrate on the simplest system which belongs to
the class of long-range random Hamiltonians, Eq. (\ref{long-range}),
with $d=1$. This is the ensemble of Hermitian critical power-law banded
random  matrices \cite{MFSeil} (CPLBRM) 
%
%
where the multifractality can be continuously tuned from 
weak to strong by changing only one parameter $0<b<\infty$:
\begin{equation}
\label{var}\langle H_{nm}\rangle=0,\;\;\;
\langle|H_{nm}|^{2}\rangle=\left\{\begin{matrix}\beta^{-1},& n=m\cr
\frac{1}{2\left[1+\left( \frac{n-m}{b} \right)^{2}\right]},&n\neq
m\end{matrix}\right.,
\end{equation}
where $\beta=1,2,4$ is the Dyson symmetry number.

We consider the limit of strong multifractality $b\ll 1$ and
demonstrate the validity of Eqs.~(\ref{Ch-cr}), (\ref{Ch-conj}),
(\ref{MF-regime}), (\ref{non-anal}) and (\ref{PL-tail}) which are the key relationships 
for the description of L\'evy flights and multifractality. Furthermore, 
we will find the function $f(x)$ explicitly for the Gaussian Orthogonal 
(GOE) and Unitary (GUE) ensembles thereby unifying both concepts 
of multifractality and L\'evy flights in the one single analytic function.

One of the purposes of this paper is to compare the density
correlations function for random critical states of a long-range
Hamiltonian with that for the classical random walks on a fractal.
For a classical random walk with a constant step length on a lattice
the density correlation function decreases exponentially at large
distances $\hat{{\cal D}}_{r,r'}(t)\propto e^{-\frac{r^{2}}{2Dt}}$
and thus all moments $\langle r^{2m}\rangle$ are well defined. A
peculiar class of random walks called {\it ``L\'evy flights''} is
realized in the models with variable step length. If the probability
to have a long step is heavily tailed, the probability distribution
of random walks may have power-law tails at large distances and thus
the moments $\langle r^{2m}\rangle$ may diverge \cite{Huges}. The
similar power-law tails may appear in classical random walk on
hierarchical manifolds \cite{Wegner-1, Wegner-long, ret-prob}, where
fractality is build in their geometry. \\
We will demonstrate that certain models of classical random walk on
a fractal result in the same regimes of the density correlation
function as for the quantum critical long-range Hamiltonians of
Eq.~(\ref{long-range}). Namely, the two principle regimes
Eqs. (\ref{MF-regime}), (\ref{PL-tail}) described in Fig.~1 have their
counterparts for the classical diffusion, with the correlation
fractal dimension $d_{2}$ being replaced by the {\it Hausdorff}
dimension $d_{h}$ and the dimensionality of space $d$ being replaced
by the {\it walk dimension}. Moreover, we will see that the critical
exponent $2/d$ of sub-diffusion $\langle r^{2} \rangle \propto
t^{2/d}$ for critical random Hamiltonians in $d>2$ dimensions
appears to have the same relationship with the exponent $-d$ of
L\'evy flights $r^{d}\,\hat{{\cal D}}_{0,r}(t)\equiv S(r,t)\propto r^{-d}$
for the quantum problem with long-range random Hamiltonians, 
Eq.~(\ref{long-range}), and for the certain classical random walks on fractals 
%
%
\cite{Wegner-1, Wegner-long, ret-prob}.
\\
The paper is organized as follows. In Section~II we review the
method of virial expansion in the number of resonant levels and give
general expressions for the density correlation function (effective
diffuson) for any almost diagonal Gaussian Random Matrix Theory (RMT). 
In Section~III, we apply
these expressions to compute the density correlation function
for the specific {\rm CPLBRM} in the coordinate/time representation
and uncover multifractality and L\'evy flights as consequences of
the long-range nature of the Hamiltonian, Eq.~(\ref{var}). In Section~IV, we
compute the density correlation function in the
coordinate/frequency representation. Sections~V and VI are devoted
to verification of the key relationships
Eqs.~(\ref{Ch-cr}), (\ref{Ch-conj}) and (\ref{non-anal}) due to
multifractality and L\'evy flights as seen in the
momentum-frequency representation of the density correlation
function. 
{\myred
We explain a perturbative derivation of L\'evy flights for strong
multifractality and the Wigner-Dyson limit of the CPLBRM in 
Sections~VII and VIII, respectively.
In Section~IX, we present numerical results for the
critical RMT with weak and strong multifractality. In Section~X, we
discuss the classical random walks on fractals and show that in
certain models they have the same phenomenology as their quantum
counterpart.
In Section~XI we summarize the main results of the paper and discuss 
their applicability to the Anderson transition in 3D systems.
}

\section{Virial expansion method}
In order to check the validity of Eqs.~(\ref{Ch-cr}), (\ref{Ch-conj}), (\ref{MF-regime})
(\ref{non-anal}) and (\ref{PL-tail}) for the ensemble of CPLBRM, Eq.~(\ref{var}), at
strong multifractality, $b\ll 1$, we exploit the method of virial
expansion in the number of resonant levels \cite{KYevOss,YevOss}.
This method is a certain {\it re-summation} of the {\it locator
expansion} \cite{Anderson} in the hopping matrix element $h_{nm}$:
\begin{equation}
\label{hop} h_{nm}^{2}=\langle |H_{n\neq
m}|^{2}\rangle=\frac{1}{2}\,\frac{b^{2}}{(n-m)^{2}}.
\end{equation}
The summation is organized so that any correlation function
$C(\kappa,\omega)$ is represented as a series in $b\ll 1$:
\begin{equation}
\label{vir-ser} 
 C(\kappa,\omega)=\omega^{\eta} \sum_{m=1}^{\infty}
    b^{m-1} C^{(m)}(\kappa L_{\omega}) \, ,
%
%
\end{equation}
where the length scale $L_{\omega}$ is itself $b$-dependent:
\begin{equation}
\label{L-w} L_{\omega}=\frac{2\sqrt{2}b}{\omega\,\beta},
\end{equation}
and $\eta$ is specific to a particular correlation function.

The advantage of this representation is that Eq.~(\ref{vir-ser}) is a
{\it functional} series, each function $C^{(m)}(\kappa L_{\omega})$
adding details of correlations which {\myred emerge from resonant 
interaction  of $m$ states. These states have energy levels 
$\varepsilon_{m}$ being anomalously close to each other within 
the interval of order $ b \ll 1$.
}
%
%
At small $b$, such ``multiple collision'' of levels
has small probability. Therefore, even the few first terms in the
series Eq.~(\ref{vir-ser}) give a very good approximation of the
correlation function.

Note that the density of states $\rho_{E}\approx
\sqrt{\frac{\beta}{2\pi}}\,e^{-\beta\,E^{2}/2}$ corresponding to
CPLBRM Eq.~(\ref{var}) has negligible variation at a scale
$E \sim b\ll 1$ {\myred [\onlinecite{OurDOS}]}. We will neglect such variations 
throughout the paper and approximate $\rho(E)\approx \rho(0)=\rho_{0}$. 
With such an accuracy we obtain for the retarded density correlation function:
%
%
\begin{equation}
\label{cc} \tilde{{\cal D}}_{r,r'}(\omega)=\frac{2\pi
\rho_{0}}{-i\omega+0}\,\left[\delta_{r,r'}+
\tilde{{\cal D}}^{(2)}_{r,r'}(\omega)+ \tilde{{\cal D}}^{(3)}_{r,r'}(\omega) 
+ \ldots \right].
\end{equation}
In Eq.~(\ref{cc}) we denote
\begin{equation}
\label{F} 2\pi\rho_{0}\tilde{{\cal
D}}_{r,r'}^{(2,3)}(\omega)=\int_{0}^{\infty}e^{i\omega
t}\,\partial_{t}\hat{{\cal D}}_{r,r'}^{(2,3)}(t)\,dt.
\end{equation}
The 
density correlation function in the time domain is given by:
\begin{equation}
\label{full-t-d}
\hat{{\cal D}}_{r,r'}(t) \simeq \left( 2\pi\rho_{0}\,\delta_{r,r'}\, + 
        \hat{{\cal D}}^{(2)}_{r,r'}(t) + \hat{{\cal D}}^{(3)}_{r,r'}(t) \right) {\myred \theta(t)} \, .
\end{equation}
{\myred To simplify formulae, we assume below $ \, t > 0 \, $ and skip $ \, \theta(t) $.}

Calculating leading terms of the virial expansion, we obtain
the following expressions for $\beta=1$ (GOE):
%
%
\begin{equation}
\label{D-t-GOE} \hat{{\cal D}}^{(2)}_{r\neq r'}(t)=2\pi
h_{r,r'}^{2}\,t\,e^{-(h_{r,r'}t)^{2}}\,I_{0}\Bigl((h_{r,r'}t)^{2}\Bigr) \, ;
\end{equation}
and for $\beta=2$ (GUE):
%
%
\begin{equation}
\label{D-t-GUE} \hat{{\cal D}}^{(2)}_{r\neq
r'}(t)=2\pi\left[h_{r,r'}^{2}\,t\,e^{-(h_{r,r'}t)^{2}}+\frac{\sqrt{\pi}}{2}\,h_{r,r'}\,{\rm
Erf}(h_{r,r'}t)\right] \, ,
\end{equation}
where $h_{r,r'}$ is defined in Eq.~(\ref{hop}); $ \, I_j (\ldots) \, $ is the 
modified Bessel function of the first kind. Answers for
subleading terms of the virial expansion are more lengthy and
we will publish them elsewhere \cite{3col}.

Equations (\ref{cc})-(\ref{D-t-GUE}) are valid for {\it any} Gaussian 
RMT with an arbitrary variance $h_{r,r'}^{2}$ of the hopping matrix 
elements that is parametrically smaller than the variance of the diagonal
ones ({\it almost diagonal Gaussian RMT}).

Remarkably,  $\hat{{\cal D}}^{(2,3)}_{r,r'}(t)$ at $r=r'$ is expressed
in terms of $\hat{{\cal D}}^{(2,3)}_{r\neq r'}(t)$ as follows:
\begin{equation}
\label{rr} \hat{{\cal D}}^{(m)}_{r,r}(t) = - \sum_{r\neq
r'}\hat{{\cal D}}^{(m)}_{r, r'}(t) \, , \ m = 2, 3.
\end{equation}
This is a consequence of the particle conservation which requires
\begin{equation}
\label{part-cons} \sum_{r}\hat{{\cal D}}_{r,r'}(t)=2\pi
\rho_{0}\, . 
\end{equation}
As this relation is already fulfilled by the first term in
Eq.~(\ref{cc}), all other terms proportional to ${\cal
D}_{r,r'}^{(m)}$ ($m\geq 2$) should obey Eq.~(\ref{rr}).

\section{Density correlation function in the time domain and return probability}
An inspection of Eqs.~(\ref{D-t-GOE}), (\ref{D-t-GUE}) shows that for
the CPLBRM with $b\ll 1$ the asymptotic behavior of the effective
diffuson $\hat{{\cal D}}_{r,r'}(t)=\rho_{0}^{-1}\,\hat{{\cal
D}}_{r,r'}^{(2)}(t)$ is given by:
\begin{equation}
\label{D-r-t-small} \hat{{\cal D}}_{r,r'}(t)= 2\pi\rho_{0}\left\{
\begin{matrix}\frac{\sqrt{2}\,b}{2|r-r'|},& {\rm GOE} \cr
 \frac{\pi\, b}{2\sqrt{2}|r-r'|},& {\rm GUE}
 \end{matrix}\right.,\;\;\;\; |r-r'|\ll b\,t \, .
\end{equation}
\begin{equation}
\label{D-r-t-large} \hat{{\cal D}}_{r,r'}(t)=
\frac{\pi\beta\,b^{2}\,t}{\,|r-r'|^{2}},\;\;\;\; |r-r'|\gg b\,t \, .
\end{equation}
In a more general case of long-range Hamiltonians obeying
Eq.~(\ref{long-range}) one should replace $|r-r'|\rightarrow
|r-r'|^{d}$ to obtain the large-distance tail Eq.~(\ref{PL-tail})
which is the hallmark of the L\'evy flights.

The small-distance behavior $\hat{{\cal D}}_{r,r'}(t)\propto
|r-r'|^{-d}$ is also remarkable. Applying Eq.~(\ref{rr}) one obtains
for the survival probability Eq.~(\ref{P-t}):
\begin{equation}
\label{surviv} P(t)=1-\frac{d_{2}}{d}\,\ln(b\,t)\approx
(b\,t)^{-\frac{d_{2}}{d}},
\end{equation}
{\myred (see Ref.[\onlinecite{Chalker-proof}] for more details)}
where for the CPLBRM one finds \cite{Mirlin-rev}:
\begin{equation}
\label{d-2} \frac{d_{2}}{d} \simeq \left\{ \begin{matrix}\sqrt{2}\,b, &
{\rm GOE}\cr \frac{\pi\,b}{\sqrt{2}}, & {\rm GUE}
\end{matrix}\right.,\;\;\;\;(b\ll 1)\,.
\end{equation}
Thus Eq.~(\ref{surviv}) shows that the asymptotic behavior
$|r-r'|^{-d}$ is a signature of (strong) multifractality with the
correlation dimension given by Eq.~(\ref{d-2}).

We conclude this section by emphasizing once again that the
multifractality and L\'evy flights for the long-range Hamiltonians
Eq.~(\ref{long-range}) have the same root and cannot exist one
without the other.

\section{Density correlation function in the frequency domain}
\begin{figure}[ht]
 \includegraphics[width=8.5cm, 
                              angle=0]{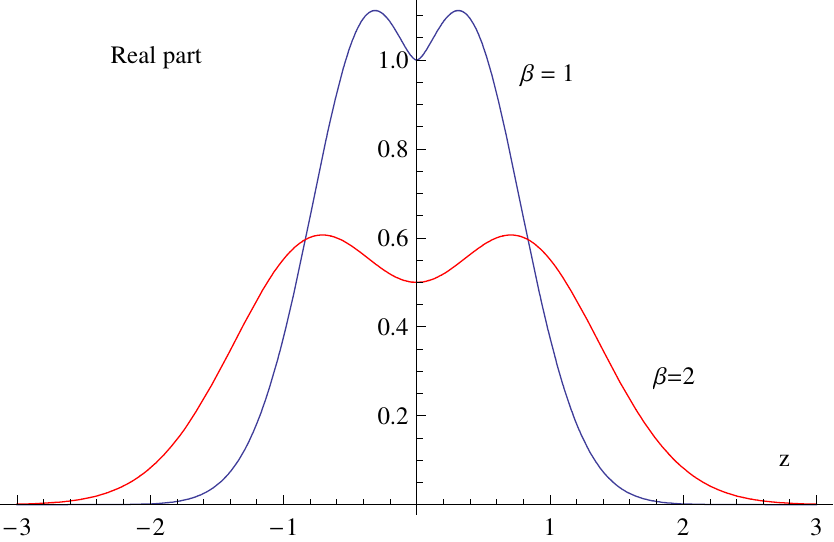}
\caption{
(color online) Real part of the functions $\Omega^{({\rm
GOE})}(z)$ in Eq.~(\ref{D-w-GOE}) (blue) and $\Omega^{({\rm
GUE})}(z)$ in Eq.~(\ref{D-w-GUE}) (red).  
{\myred
They describe the
$\omega$-dependence of the corrections $\tilde{{\cal D}}^{(2)}_{r,r'}$ 
to the retarded density correlation function at a fixed $r\neq r'$.
Note that these corrections are caused by the extended nature 
of wave functions. 
}
%
%
}
\label{Re-w}
\end{figure}

\begin{figure}[ht]
 \includegraphics[width=8.5cm, 
                              angle=0]{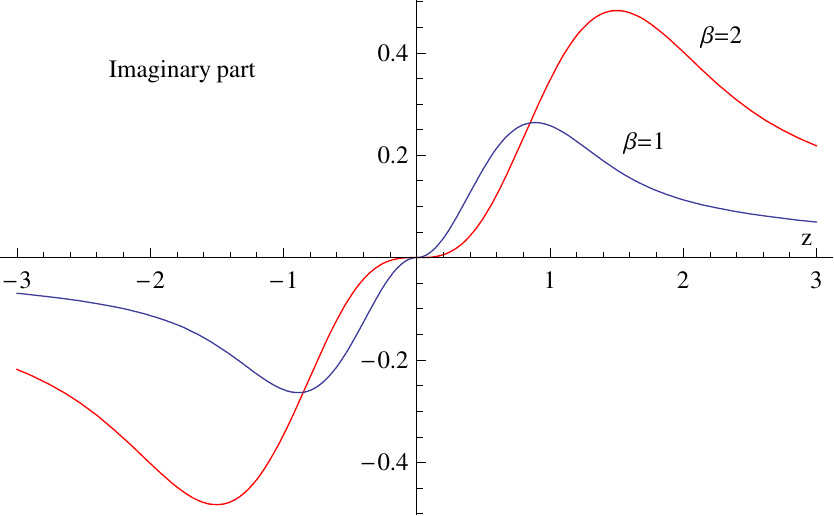}
\caption{(color online)  Imaginary part of the functions
$\Omega^{({\rm GOE})}(z)$ in Eq.~(\ref{D-w-GOE}) (blue) and
$\Omega^{({\rm GUE})}(z)$ in Eq.~(\ref{D-w-GUE}) (red).} \label{Im-w}
\end{figure}

One can invert Eq.~(\ref{F}) and obtain:
\begin{equation}
\label{def-rw} \tilde{{\cal D}}_{r\neq
r'}(\omega)=\frac{\sqrt{2\pi}\beta^{\frac{3}{2}}\pi^{\beta-1}}{4}\,\frac{i}{z}\,\Omega^{(\beta)}(z),
\end{equation}
where
\begin{equation}
\label{z} z=(|r-r'|/L_{\omega})^{d}\propto \omega
\,|r-r'|^{d} \, ,
\end{equation}
$ d=1 \, $ for CPLBRM, $L_{\omega}$ is given by Eq.~(\ref{L-w})
and expressions for $ \, \Omega \, $ read: 
%
%
\begin{eqnarray}
\label{D-w-GOE} && \Omega^{({\rm GOE})}(z)=z^{2}e^{-z^{2}}\,\left[
\left(K_{0}(z^{2})+K_{1}(z^{2})\right)\right.+\nonumber \\
&&\left.\pi i\,{\rm sgn}(z)\,\left(I_{0}(z^{2})-I_{1}(z^{2})
\right)\right],\;\;\;({\rm GOE}, \beta = 1) \, ;
\end{eqnarray}
%
%
\begin{eqnarray}
\label{D-w-GUE} && \Omega^{({\rm
GUE})}(z)=(z^{2}+\frac{1}{2})e^{-z^{2}} \times \\
&&\left[1+ i\,\left( {\rm
Erfi}(z)-\frac{z\,e^{z^{2}}}{\sqrt{\pi}\,(z^{2}+\frac{1}{2})}\right)\right],\;\;\;({\rm
GUE}, \beta = 2) \, ;
\nonumber
\end{eqnarray}
$ \, K_j (\ldots) \, $ is the modified Bessel function of the second kind.
Functions $\Omega^{(\beta)}(z)$ in Eqs.~(\ref{D-w-GOE}), (\ref{D-w-GUE}) 
are qualitatively very similar for the orthogonal and the unitary
ensemble (see Figs.~\ref{Re-w},\ref{Im-w}).
Their real parts are even functions of $z$ which exponentially
decrease like $|z|^{\beta}\,e^{-2z^{2}/\beta}$ at $|z|\gg 1$. The
imaginary parts are odd functions of $z$ which behave like $z^{-1}$
at large $|z|$. Correspondingly, the density correlation function in
the coordinate-frequency representation behaves at $|z|\gg 1$ (large
distances or frequencies) like:

\begin{eqnarray}
\label{asympt-Re} {\rm Im}[\tilde{{\cal D}}_{r\neq r'}(\omega)] & \sim &
{\rm sign}(z)\,|z|^{\beta-1}\,e^{-2z^{2}/\beta}, \\
\label{asympt-Im} {\rm Re}[\tilde{{\cal D}}_{r\neq r'}(\omega)] & \sim &
-\frac{1}{z^{2}}.
\end{eqnarray}
Note that the correlation function $\tilde{{\cal D}}_{r\neq r'}(\omega)$ for
$|r-r'|\gg L_{\omega}$ has a power-law tail. This is another
manifestation of L\'evy flights in the $r-\omega$ domain.


\section{Frequency-momentum representation.}

Now we compute the retarded density correlation function ${\cal
D}(\kappa,\omega)$ in the momentum-frequency representation. This
allows us to make contact with Eq.~(\ref{Ch-cr}) to check Chalker's
conjecture on the effect of multifractality Eq.~(\ref{Ch-conj}) and
to detect the non-analytic in $\kappa$ terms in Eq.~(\ref{non-anal}),
which is a signature of L\'evy flights.

To this end we define:
\begin{equation}
\label{def-kw} {\cal
D}^{(2)}(\kappa,\omega)=\frac{1}{N}\sum_{r,r'}\tilde{{\cal
D}}_{r,r'}^{(2)}(\omega)\,e^{i\kappa\,(r-r')},
\end{equation}
where $N\times N$ is the size of the matrix. We assume this size to
be larger than all the relevant length scales of the system. If in
addition $\kappa\ll 1$, one can switch from the sum to integral in
Eq.~(\ref{def-kw}) to implement the {\it continuous approximation}
which is valid at $|L_{\omega}|\gg 1$.

Before proceeding with calculations, let us note that Eq.~(\ref{rr})
implies:
\begin{equation}
\label{k0} {\cal D}^{(2)}(\kappa=0,\omega)=0.
\end{equation}
Thus one can do the Fourier transform of $\tilde{{\cal
D}}^{(2)}_{r,r'}(\omega)$ at $|r-r'|\neq 0$ and then subtract the
$\kappa=0$ value of the result obtained. Thus, we obtain from
Eqs.~(\ref{D-w-GOE}), (\ref{D-w-GUE}):
\begin{equation}
\label{def-Lw} {\cal
D}(\kappa,\omega)=\frac{2\pi\rho_{0}}{-i\omega+0}+|L_{\omega}|\,G(\bar{\kappa}),
\end{equation}
where $\bar{\kappa}=\kappa\,L_{\omega}$. 
One can easily see that Eq.~(\ref{def-Lw}) is consistent with Eq.~(\ref{Ch-cr}) in the 
limit $ \, b \log (\bar{k}) \ll 1 $, with  $ \, f(x) = i (\sqrt{2}/\pi  \beta \rho_0) \, b \, G(\bar{k}) $.
%
%
The retarded character of the correlation function is encoded in real and
imaginary parts of $ \, G(\bar{k}) \, $ which obey the Kramers-Kronig relationship.
{\myred The expressions for them read
}

\vspace{0.25 cm}
{\myred
GOE, $ \, \beta = 1$:
}
\begin{equation}
\label{Re-kw-GOE} {\rm Re}[G^{({\rm
GOE})}(\bar{\kappa})]=\frac{(2\pi)^{2}}{32}\,
|\bar{\kappa}|\,e^{-\frac{\bar{\kappa}^{2}}{16}}\,I_{0}\left(
\frac{\bar{\kappa}^{2}}{16}\right),
\end{equation}
\begin{eqnarray}
\label{Im-kw-GOE} &&{\rm Im}[G^{({\rm GOE})}(\bar{\kappa})]=-{\rm
sign(\omega)}\frac{\sqrt{2\pi}}{4}\times \nonumber
\\ &&\left[ 1 + \frac{\bar{\kappa}^{2}}{3}\,{\rm pFq}\left(\{ 1,1,3\},\left\{
\frac{3}{2},2,\frac{5}{2}\right\},-\frac{\bar{\kappa}^{2}}{8}
\right)\right.-\nonumber
\\ &&\left.
{\rm pFq}\left(\{ 1,1\},\left\{\frac{1}{2},\frac{3}{2} \right\},
-\frac{\bar{\kappa}^{2}}{8} \right)\right],
\end{eqnarray}
{\myred

GUE, $ \, \beta = 2$:
}
%
%
\begin{equation}
\label{Re-kw-GUE} {\rm Re}[G^{({\rm
GUE})}(\bar{\kappa})]=\frac{\pi^{2}}{2}\,\left[
|\bar{\kappa}|\,e^{-\frac{\bar{\kappa}^{2}}{4}}-\sqrt{\pi} \,{\rm
Erf}\left( \frac{\bar{\kappa}}{2}\right) \right].
\end{equation}
\begin{eqnarray}
\label{Im-kw-GUE} {\rm Im}[G^{({\rm GUE})}(\bar{\kappa})]&=&-{\rm
sign(\omega)}\frac{\pi^{\frac{3}{2}}}{4}\,\left[
2\sqrt{\pi}\,e^{-\frac{\bar{\kappa}^{2}}{4}}\,\bar{\kappa}\, {\rm
Erfi}\left( \frac{\bar{\kappa}}{2}\right)+\right.\nonumber
\\ &&\left.\bar{\kappa}^{2}\,{\rm pFq}\left(\{ 1,1\},\left\{
\frac{3}{2},2\right\},-\frac{\bar{\kappa}^{2}}{4}\right)\right] \, .
\end{eqnarray}
Here $\gamma$ is the Euler constant and  ${\rm pFq}(...)$ is the
hypergeometric function. We discuss these results in the next section.

\section{Comparison with the Chalker's conjecture.}
The asymptotic behavior of the functions $G^{{\rm
GOE}}(\bar{\kappa})$ and $G^{{\rm GUE}}(\bar{\kappa})$ is very
similar. At small $\bar{\kappa}\ll 1$:
\begin{equation}
\label{as-re-small} {\rm Re}[G(\bar{\kappa})]=\left \{
\begin{matrix} \frac{1}{2}\pi^{2}\,|\bar{\kappa}|, &
 \beta=1 \cr
\pi^{2}\,|\bar{\kappa}|, & \beta=2\end{matrix}\right.\,,
\end{equation}
\begin{equation}
\label{as-im-small} {\rm Im}[G(\bar{\kappa})]=\left \{
\begin{matrix} -\frac{1}{4}\sqrt{\frac{\pi}{2}}\,\bar{\kappa}^{2}, &
 \beta=1 \cr
-\frac{3}{8}\,\bar{\kappa}^{2}, & \beta=2\end{matrix}\right. \,.
\end{equation}
This is consistent with Eq.~(\ref{non-anal}) (at $d=1$) where all
$c_{i}\sim b$. In particular, the non-analytic behavior $\propto
|\bar{\kappa}|$ caused by the L\'evy-flights is confirmed by
Eq.~(\ref{as-re-small}). 
{\myred Eq.~(\ref{as-im-small}) corresponds
to corrections $\propto c_{2}\,x^{2}$ in Eq.~(\ref{non-anal}).
}
%
%

\begin{figure}[ht]
\includegraphics[width=8.5cm, 
                            angle=0]{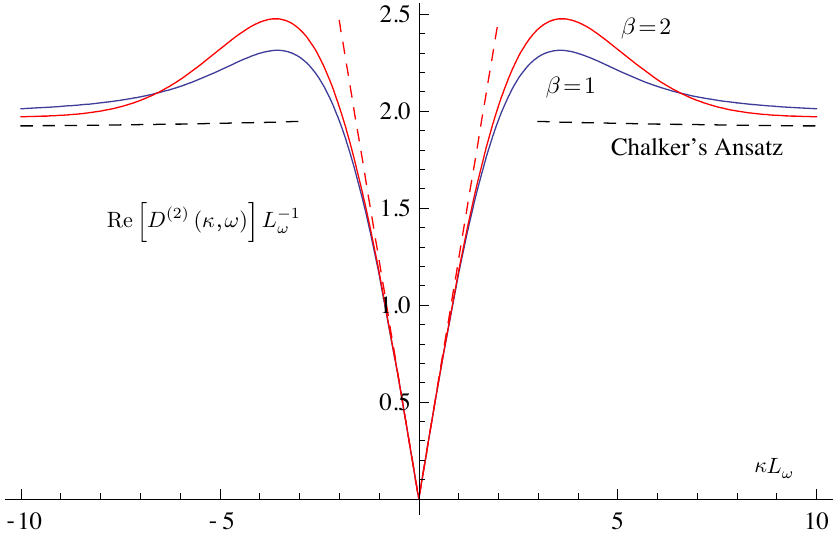}
\caption{(color online) Real part of $G(\bar{k})$ for ${\rm GOE}$
(blue) and {\rm GUE} (red). The dashed lines is the asymptotic
behavior given by Eq.~(\ref{non-anal}) (red) and Eq.~(\ref{Ch-conj})
(black) at $d_{2}=0.01$ and $c=1+O(b)$. For better comparison the
{\rm GUE} curve is given in the re-scaled coordinates so that the
asymptotic behavior at small and large $\kappa$ coincides with that
of the {\rm GOE} curve. } \label{Re-D-kw}
\end{figure}

More interesting is the asymptotic behavior {\myred of $G(\bar{k})$} 
at large $|\bar{\kappa}|\gg 1$. 
%
%
While ${\rm Re}[G(\bar{\kappa})]$ tends to a constant
\begin{equation}
\label{as-re-large} {\rm Re}[G(\bar{\kappa})]=\left \{
\begin{matrix} \frac{\pi^{\frac{3}{2}}}{2\sqrt{2}}, &
 \beta=1 \cr
\frac{\pi^{\frac{5}{2}}}{2}, & \beta=2 \end{matrix}
\right.\,\,\,\;\;\;(|\bar{\kappa}|\gg 1),
\end{equation}
$|{\rm Im}[G(\bar{\kappa})]|$ increases logarithmically

\begin{equation}
\label{as-im-large} {\rm Im}[G(\bar{\kappa})]=\left \{
\begin{matrix} -\frac{\sqrt{2\pi}}{2}\,\ln(|\bar{\kappa}|)+{\rm const}, &
 \beta=1 \cr
-\frac{1}{2}\,\ln(|\bar{\kappa}|)+{\rm const}, &
\beta=2\end{matrix}\right.\,.
\end{equation}
Such a behavior is fully consistent with Eq.~(\ref{Ch-conj}), in which a sub-leading
term is added:
\begin{equation}\label{Ch-conj-Corr}
  f(x) =c x^{d_{2}} - c \, .
\end{equation}
For very large $ \, x \gg \exp(1/b) $, Eq.~(\ref{Ch-conj-Corr}) reduces to Eq.~(7) 
while in the intermediate region $ \, 1 \ll  x \ll \exp(1/b) $, which is present only 
at small $b\ll 1$, one obtains:
\[
   f(x) \simeq c d_{2} \log( x ) \, .
\]
%
%
Thus we confirm the Chalker's conjecture for the density
correlation function to the first order in the small parameter $b \ll 1$.

\begin{figure}[ht]
\includegraphics[width=8.5cm, 
                             angle=0]{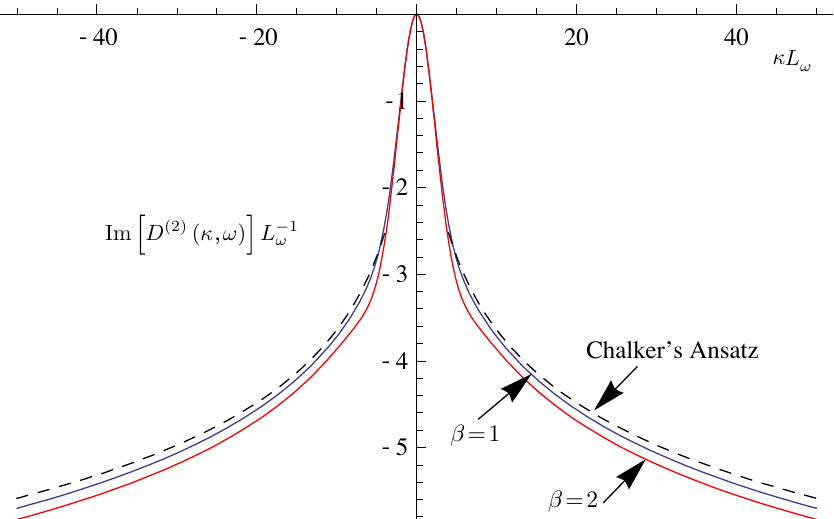}
\caption{(color online) Imaginary part of $G(\bar{k})$ for {\rm GOE}
(blue) and {\rm GUE} (red). The latter curve is given in the
re-scaled coordinates as in Fig.~\ref{Re-D-kw}. The black dashed line
represents the asymptotic behavior Eq.~(\ref{Ch-conj}) for the proper
choice of the coefficient $c=1+0.90\,b$ and $d_{2}=0.01$.}
\label{Im-D-kw}
\end{figure}

In order to illustrate the similarity of the density correlation
functions conjectured in Eqs.~(\ref{Ch-cr}), (\ref{Ch-conj}) and
(\ref{non-anal}) and obtained analytically for the CPLBRM
Eq.~(\ref{var}), we plot in Fig.~\ref{Re-D-kw} and Fig.~\ref{Im-D-kw}
the real and imaginary parts of $G(\bar{\kappa})$ together with the
asymptotic behavior Eq.~(\ref{Ch-conj}) at a small $d_{2}=0.01$. The
choice of only one constant, $c=1+O(b)$, ensures both the right
leading logarithmic behavior of ${\rm Im}[G(\bar{\kappa})]$ and the
correct limit of ${\rm Re}[G(\bar{\kappa})]$ as 
{\myred $\bar{\kappa}\rightarrow \infty$}.   
It is also seen that upon a proper re-scaling of $x$ and
$y$ coordinates the shape of $G^{{\rm GOE}}(\bar{\kappa})$ and
$G^{{\rm GUE}}(\bar{\kappa})$ is very similar. We conclude
therefore that the shape of the $G(\bar{\kappa})$ dependence is
dictated mostly by criticality encoded in the Chalker's conjecture
and not by the symmetry of the Hamiltonian.
\section{L\'evy flights for strong multifractality: a poor-men derivation.}
In this section, we exploit another expression for the density correlation function 
(at a fixed energy of the wave packet $\varepsilon$ {\myred which has been 
reinstated in this Section}) 
in the time domain which is equivalent to Eq.~(\ref{D-r}) provided that $\omega\ll B$ 
or $t\gg 1/B$, where $B$ is the energy bandwidth:
\begin{eqnarray}
\label{t-sum}
\hat{{\cal D}}_{r,r'}(t)&=&2\pi\sum_{n,m}\langle \delta(\varepsilon-\varepsilon_{n})
e^{it(\varepsilon_{m}-\varepsilon_{n})}\\ \nonumber & &\psi_{n}(r)\psi_{n}^{*}(r+R)\psi_{m}(r+R)\psi_{m}^{*}
(r)\rangle.
\end{eqnarray}

Note that  $\hat{{\cal D}}_{r\neq r'}(t=0)=0$ due to completeness of the set of normalized wave functions. 
Therefore,
one can replace $e^{it(\varepsilon_{n}-\varepsilon_{m})}$
by $e^{it(\varepsilon_{n}-\varepsilon_{m})}-1$ without changing the result.

Let us consider very strong multifractality and very large distances $R=r-r'$. Then the wave function 
can be approximated as
\begin{equation}
\label{psi-appr}
\psi_{n}(r\neq n)\approx \frac{H_{n,r}}{\varepsilon_{n}-\varepsilon_{r}},\;\;\;\;\psi_{n}(r=n)\approx1,
\end{equation}
where $H_{nm}$ is an off-diagonal entry of the matrix Hamiltonian and $\varepsilon_{n} \equiv H_{nn}$ is
the diagonal entry. In the case of very small $b$, when the wave functions normalization is almost
all concentrated at one site $n$ (center of localization), the eigenenergy of a state $\varepsilon_{n}$
is almost equal to the on-site energy $H_{nn}$.

The main contribution to Eq.~(\ref{t-sum}) is given by two terms
$(m=r, n=r+R)$ or $(n=r, m=r+R)$. In both cases the combination of wave functions is equal to:
$$
-\frac{|H_{r,r+R}|^{2}}{(\varepsilon_{r}-\varepsilon_{r+R})^{2}} \, .
$$
{\myred
Now, we average over disorder. For a Gaussian ensemble, averaging over 
$\varepsilon_{n} = H_{nn}$ and over the hopping matrix elements $H_{nm}$ are 
independent  from each other. Therefore, we replace $|H_{nm}|^{2}$ by its average
$\langle |H_{nm}|^{2}\rangle$ given by Eq.~(\ref{var}) and reduce averaging over
$ \varepsilon_{n} $ to the energy integral with the help of the spectral correlation 
function $R(\varepsilon, \varepsilon')= \langle \rho(\varepsilon)\,\rho(\varepsilon')\rangle$:
}
%
%
\begin{equation}
\label{dcf-appr}
\hat{{\cal D}}_{r,r+R}(t)\approx \int_{-\infty}^{+\infty} d\varepsilon'\,R(\varepsilon,\varepsilon')
\, \frac{1-e^{it(\varepsilon-\varepsilon')}}{(\varepsilon-\varepsilon')^{2}}
\,\frac{2\pi\,b^{2}}{R^{2}}\,.
\end{equation}
Note that due to {\it level repulsion} $R(\varepsilon, \varepsilon)=0$
the integral in Eq.~(\ref{dcf-appr}) is convergent at
$\varepsilon'=\varepsilon$. However, at $ \, t \, $ {\myred being smaller than the Heisenberg time,
$t\ll t_{{\rm H}} = N\rho_{0}$}
(or $\omega\gg \Delta\sim B/N$), the region of level repulsion is very narrow so that one can
replace $R(\varepsilon, \varepsilon')\approx \rho(\varepsilon)\rho(\varepsilon')$, where
$\rho(\varepsilon)$ is the averaged DoS.
Using also the symmetric form of the averaged DoS $\rho(\varepsilon)=\rho(-\varepsilon)$ we obtain at
$\varepsilon=0$:
\begin{eqnarray}
\label{E-eq-zero}
\hat{{\cal D}}_{r,r+R}(t)\approx \frac{4\pi\rho_{0}\,b^{2}}{R^{2}}
\int_{0}^{+\infty} d\varepsilon'\,\rho(\varepsilon')
\, \frac{1-\cos(t\varepsilon')}{\varepsilon'^{2}}.
\end{eqnarray}
At $t\gg 1/B$, the integral in Eq.~(\ref{E-eq-zero}) is dominated by $\varepsilon'\ll B$
%
%
and one can replace $\rho(\varepsilon')\approx \rho_{0}$ and obtain:
%
%
\begin{equation}
\label{lin-te}
\hat{{\cal D}}_{r,r+R}(t)\approx \frac{4\pi\rho_{0}^{2}\,b^{2}}{R^{2}}\,t
\int_{0}^{+\infty} dx
\, \frac{1-\cos(x)}{x^{2}}=\frac{2\pi^{2}\rho_{0}^{2}\,b^{2}}{R^{2}}\,t \,.
\end{equation}
This is exactly the result Eq.~(\ref{D-r-t-large}) obtained in the corresponding
limit $R\gg bt$ from general formulae of the virial expansion.

Note, however, that 
Eq.~(\ref{lin-te}) does not describe the behavior of the correlation
function Eq.~(\ref{t-sum}) at smallest $t\ll \rho_{0}\sim 1/B$. In this case, one 
can expand the exponent in Eq.~(\ref{E-eq-zero}) and arrive at:
%
%
\begin{equation}
\label{kvadr-te}
\hat{{\cal D}}_{r,r+R}(t)\approx \frac{2\pi\rho_{0}\,b^{2}}{R^{2}}\,t^{2}
\int_{0}^{+\infty} dx \rho(x)=\frac{\pi\rho_{0}b^{2}}{R^{2}}\,t^{2}.
\end{equation}
The two asymptotics Eq.~(\ref{lin-te}) and Eq.~(\ref{kvadr-te}) match each other at $t\sim\rho_{0}$.
They both describe the L\'evy flights. However, Eq.~(\ref{lin-te}) has the {\it scaling form}
$R^{-d}\,(t/R^{-d})$ (at $d=1$) while Eq.~(\ref{kvadr-te}) is the fully perturbative result which violates
the critical scaling law $\hat{{\cal D}}_{r,r+R}(t)=R^{-d}\,S(R^{d}/t)$.
In the {\it macroscopic thermodynamic limit}, when
$R\rightarrow\infty$, $t/R^{d}$ fixed but $R/L\rightarrow 0$, the
perturbative region plotted as a function $t/R^{d}$ shrinks to zero.

\section{Density correlation function in the Wigner-Dyson limit $ \, b \to \infty $}
%
%

One can consider another limit when the size of the system $L=N$ is fixed and the parameter
$b$ in Eq.~(\ref{var}) 
is increasing. The limit $b\rightarrow\infty$ corresponds to the Wigner-Dyson RMT. 
{\myred 
In this limit, the system becomes effectively zero dimensional and 
is not critical any longer. Nevertheless, we derive and briefly review its 
density correlation function for the sake of completeness.
}
%
%

We start with the same Eq.~(\ref{t-sum}) as in the previous section but employ the independent averaging over
eigenvalues and eigenfunctions. The former is given by the famous Wigner-Dyson statistics \cite{Mehta}
while the latter is described by the Porter-Thomas statistics. The simplest case is $\beta=2$ when
the eigenfunction $\psi_{n}(r)\equiv U_{nr}$ is uniformly distributed over the unitary group ${\cal U}(N)$.
The averages of the products of $\psi$-functions are well known, e.g.:
\begin{eqnarray}
\label{prod}
&&\langle U_{n r_{1}}U_{n' r_{2}}^{*} U_{m r_{2}'}U_{m' r_{1}'}^{*} \rangle =
\frac{1}{N^{2}-1}\\ \nonumber &&(\delta_{n n'}\delta_{r_{1}r_{2}}\delta_{m m'}\delta_{r_{1}'r_{2}'}+
\delta_{n m'}\delta_{r_{1}r_{1}'}\delta_{m n'}\delta_{r_{2}r_{2}'})-\\ \nonumber
&& \frac{1}{N(N^{2}-1)}(\delta_{n n'}\delta_{r_{1}r_{1}'}\delta_{m m'}\delta_{r_{2}'r_{2}}+
\delta_{n m'}\delta_{r_{1}r_{2}}\delta_{m n'}\delta_{r_{2}'r_{1}'}).
\end{eqnarray}
Using Eq.~(\ref{prod}), we obtain:
\begin{equation}
\label{res-WD}
(2\pi\rho_{0})^{-1}\,\hat{{\cal D}}_{r,r+R}(t)= \delta_{R,0}+ (K(t)-N)\,
\frac{\delta_{R,0}N-1}{N^{2}-1},
\end{equation}
where $K(t)$ is the {\it spectral form-factor}:
\begin{equation}
\label{K-ff}
K(t)=\frac{1}{N\rho_{0}}\,\sum_{n,m}\langle \delta(\varepsilon-\varepsilon_{n})\,
e^{it(\varepsilon_{m}-\varepsilon_{m})} \rangle. 
\end{equation}
At $t=0$, summations over $n$ and $m$ are independent and one finds $K(t=0)=N$
while at large $t\rightarrow\infty$ only terms with $n=m$ survive and $K(t\rightarrow\infty)=1$.
However, the behavior of $K(t)$ at small times is highly uneven: $K(t)$ drops to almost zero for a
very short time of the order of the inverse bandwidth $\rho_{0}\sim 1/B\propto 1/\sqrt{N}$ and
then it recovers to unity; for example, in the unitary ensemble \cite{Mehta}: $ \, K = t / t_{{\rm H}} \, $ 
for $\rho_{0}\ll t<t_{{\rm H}} $ and $ \, K = 1 $ for $ \,t >  t_{{\rm H}}; $ {\myred we remind that the 
Heisenberg time is $  t_{{\rm H}} = N\rho_{0} $.}

%
%

One can immediately see that at $t=0$ Eq.~(\ref{res-WD}) is the {\myred discrete} $\delta$-function $\delta_{R,0}$ 
and that the sum rule Eq.~(\ref{sum-rule}) is fulfilled at any $t$. Furthermore, the density correlation function 
remains a combination of a $\delta$-function and a flat background at any $t$. The weight of the $\delta$-function
determines the return probability:
\begin{equation}
\label{return-WD}
P(t)=1-\frac{N}{N^{2}-1}\,(N-K(t)).
\end{equation}
The non-monotonic behavior of $K(t)$ results in the similar behavior of the return probability: for a short
time $t\sim \rho_{0}\sim 1/\sqrt{N}$, the wave packet leaves the origin almost completely. However, at
later times it accumulates again in the origin reaching the value $P(t)=\frac{1}{N+1}$ at
$t>t_{{\rm H}}$. Such an {\it echo} behavior is typical for a chaotic quantum system of finite size.

The flat background behaves with time as
\begin{equation}
\label{flat}
2\pi\rho_{0}\,\hat{{\cal D}}_{r,r+R}(t)=\frac{N-K(t)}{N^{2}-1}, \;\;\;\;(R>0).
\end{equation}
It rapidly grows at small times, then decreases a bit and reaches the constant
limit $1/(N+1)$ at $t>t_{{\rm H}}$.

Despite all unevenness of the behavior discussed above, there is one rough feature: at finite $R$ the
flat background is always small like $N^{-1}$. Given that the critical Hamiltonian Eq~(\ref{long-range})
approaches the Wigner-Dyson RMT at $b\rightarrow N$, one expects the density correlation function to decrease
at constant (large) system size $N$ as the parameter $b>1$ increases.

\section{Numerical results}

\begin{figure}[ht]
 \includegraphics[width=8.5cm, 
                              angle=0]{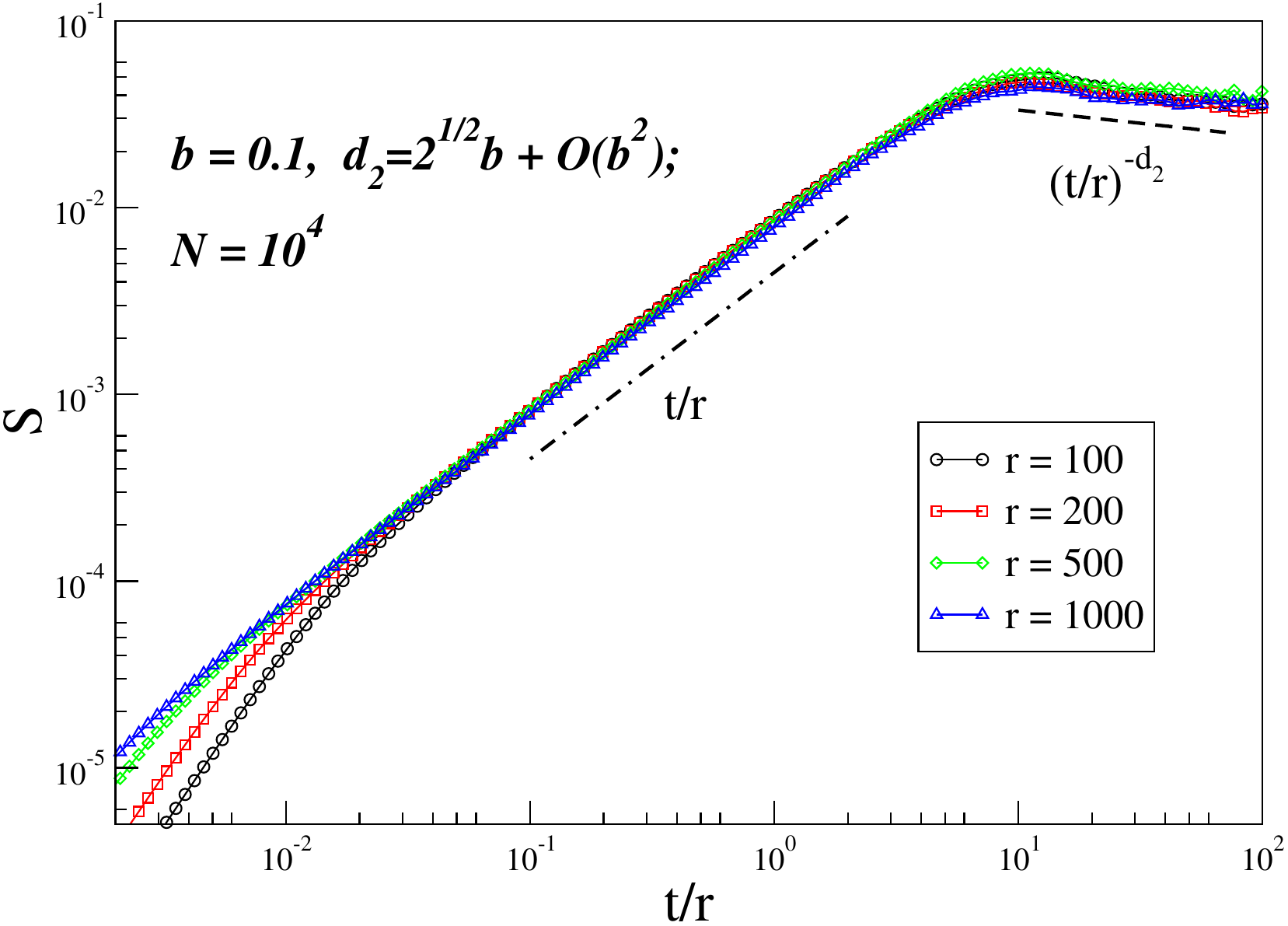}
\caption{(color online) Scaling function {\myred $S = r\,\hat{{\cal D}}_{r',r'+r}(t)$}
%
%
averaged over $r'$
for different fixed $r=100, 200,500,1000$ for the long-range
critical random matrix ensemble 
{\myred Eq.~(\ref{var}) with $ \, \beta = 1, \ b = 0.1 \, $}
and the matrix size $N=10^{4}$. The curves for the multifractal
regime Eq.~(\ref{MF-regime})
and for the L\'evy flights Eq.~(\ref{PL-tail}) collapse to the one single scaling curve. The perturbative L\'evy
flights Eq.~(\ref{kvadr-te})
at the smallest $t$ do not obey the scaling.} \label{small-b}
\end{figure}

\begin{figure}[ht]
 \includegraphics[width=8.5cm, 
                              angle=0]{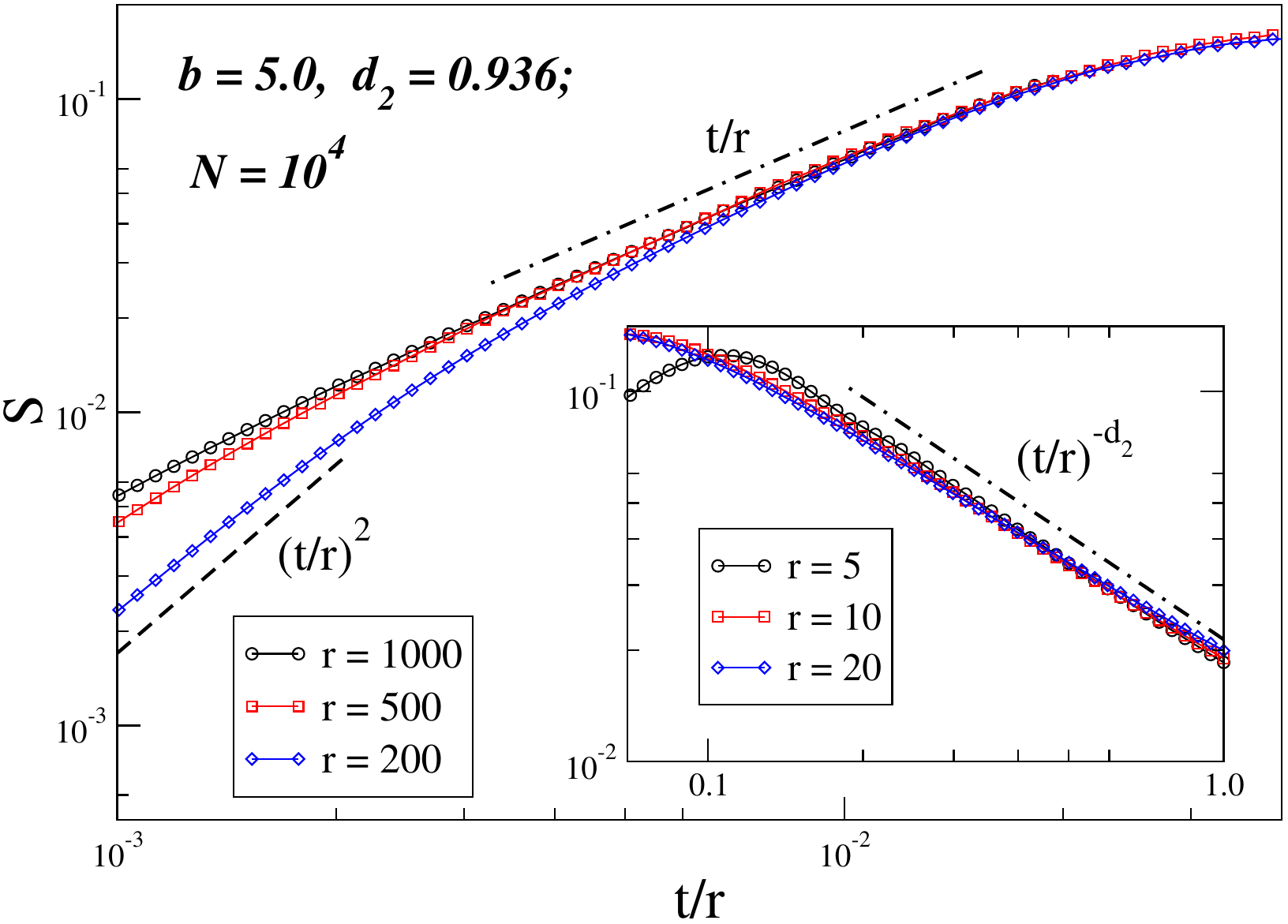}
\caption{(color online) 
 {\myred 
Scaling function $S = r\,\hat{{\cal D}}_{r',r'+r}(t)$ averaged over $r'$
for different fixed $r$ for the long-range critical random matrix ensemble 
{\myred Eq.~(\ref{var}) with $ \, \beta = 1, \ b = 5 \, $}
and the matrix size $N=10^{4}$.
{\bf Main plot}: $r=200,500,1000$. The region of
L\'evy flights with scaling Eq.~(\ref{PL-tail}) shrinks compared to the case of small $b=0.1$. The 
region of perturbative L\'evy flights Eq.~(\ref{kvadr-te}) increases with decreasing $r$.
{\bf Inset}:  the same scaling function for small
$r=5,10,20$ and larger values of $t/r$. The multifractal regime, Eq.~(\ref{MF-regime}), is well seen 
and it crosses over directly to the perturbative L\'evy flights, Eq.~(\ref{kvadr-te}), at small $t/r$.
}
%
%
} \label{large-b}
\end{figure}

In order to check the analytical results of the previous sections and, more importantly, in order
to verify their robustness for long-range critical random matrix models with weak multifractality,
we performed statistical analysis of eigenvalues and eigenfunctions obtained by direct diagonalization of
large matrices drawn from the Gaussian 
{\myred orthogonal ensemble Eq.~(\ref{var}) with $ \, \beta = 1$.}
The results are summarized in Fig.~\ref{small-b} and Fig.~\ref{large-b}.

In Fig.~\ref{small-b}, we demonstrate the scaling Eq.~(\ref{scaling})
in the two
main scaling regimes Eqs.~(\ref{MF-regime}), (\ref{PL-tail}) for the case of {\it strong multifractality} $b=0.1$.
Both the multifractal regime of Chalker and the L\'evy flights $\propto t/r^{2}$ are well seen and the
scaling Eq.~(\ref{scaling}) is confirmed by a collapse of curves for several values of the distances $r$ to
the one single
scaling curve. The scaling Eq.~(\ref{scaling}) is violated at very
small $t<\rho_{0}$ where inhomogeneity of spectrum (finite size effects in the energy space)
should be taken into account and Eq.~(\ref{kvadr-te}) stands for Eq.~(\ref{PL-tail}).
Another possible source of violation of scaling is the finite-size effects in the coordinate space
which exhibit themselves as the corrections in the parameter $r^{d}/N\ll 1$ and are important at not very small
$r^{d}/N$.

In Fig.~\ref{large-b}, we show the critical density correlation function for {\it weak
multifractality} in the case $b>1$. Besides the scaling and the Chalker's ansatz for the power-law
behavior Eq.~(\ref{MF-regime}) in the multifractal region, the focus of our study was the power-law tail
at small $t/r$ that describes the L\'evy flights. It is seen that the region of the ``scaling" L\'evy flights
Eq.~(\ref{PL-tail}) is much more narrow for $b>1$ compared to the case of small $b$ and is shrinking as
$r$ decreases. However, the exponent $p$ of the power-law in the scaling function
$S(x)\propto x^{-p}$, ($x\gg 1$)
stays constant $p=1$ and independent of the strength of multifractality, {\myred i.e. of the parameter} $b$.
This result is not obvious for large $b$ where the arguments behind the derivation of Eq.~(\ref{lin-te})
no longer apply. It was also verified analytically by considering the next order in the virial expansion 
\cite{3col}. In the next section, we will consider the classical analogue of such a behavior.

\section{Random walks over hierarchical manifold with level-dependent asymmetric rate}
\begin{figure}[ht]
\includegraphics[width=8.5cm, 
                             angle=0]{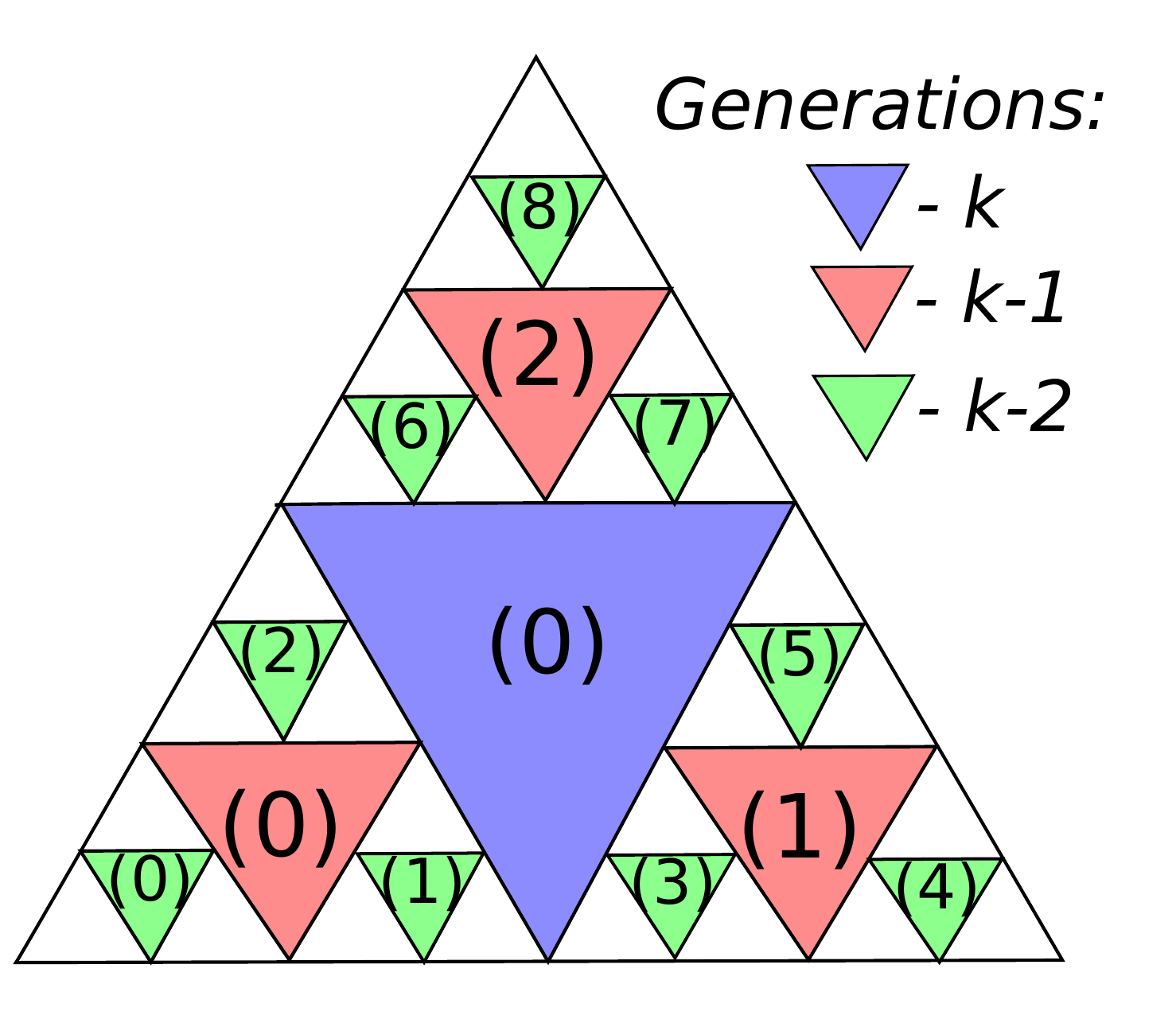}
\caption{(color online) The example of a 2d hierarchical manifold:
jumps occur only between the full {\myred (colored)} triangles of the neighboring generations (levels)
shown by different colors. There are neither jumps between triangles of the same level (size),
nor jumps between triangles that are not directly touching each other.} 
\label{gasket}
\end{figure}

In Ref.[\onlinecite{Wegner-1}] Wegner and Grossmann (WG) suggested a
classical random walk process on an hierarchical structure that is
rich enough to mimic many relevant regimes in transport in
disordered systems and in fluid turbulence. Its rigorous definition
is given in the original work. Here we illustrate this process for
the 2d Sierpinski gasket, see Fig.~\ref{gasket}. As any regular
fractal, the Sierpinski gasket is characterized by the hierarchy of
self-similar clusters, each one containing $z$ clusters of the next
generation (level). For the Sierpinski gasket of Fig.~\ref{gasket}
$z=3$. Another important parameter of the fractal geometry is the
space scaling factor $\mu>1$ that is the ratio of linear sizes of
the clusters of the $k$-th and the $k+1$-th levels. For the
Sierpinski gasket of Fig.~\ref{gasket} $\mu=2$. As in the original
work [\onlinecite{Wegner-1}], we assume that there is a smallest level $k=0$
with the size of triangle equal to 1. Then, the size of the triangle
of the largest level $k=l$ is $L=\mu^{l}$ and there are $z^{l}$ white
triangles of the smallest level in it. The parameters $z$ and $\mu$
determine the Hausdorff fractal dimension, i.e. the exponent $d_{h}$
that governs the scaling $S\propto L^{d_{h}}$ of the total area of
white triangles as the size of the largest triangle $L$ increases.
Given that $S\propto z^{l}$, one immediately finds that
\begin{equation}
\label{Haus} d_{h}=\frac{\ln z}{\ln \mu}.
\end{equation}

The WG random walk process can be viewed as jumps over the manifolds
of centers of the full  {\myred (colored)} triangles each of them being
determined by two numbers $(k,i)$, where $k$ is the level of the
triangle and $i$ enumerates in a certain way (see Fig.~\ref{gasket})
all the $z^{k-l}$ triangles of the given level $k$. An important
rule is that a jump may occur only by one level up and by one level
down. The up-jump may occur from each of $z$ smaller full triangles
of the $k-1$ level to the one single larger triangle of the $k$-th
level which smaller triangles touch directly. It happens with the
rate:
\begin{equation}
\label{prob-up} W_{k}^{({\rm up})}=s^{k}.
\end{equation}
The opposite process of down-jump is supposed to have a rate:
\begin{equation}
\label{prob-down} W_{k}^{({\rm down})}=\frac{s^{k}}{z\,r}.
\end{equation}
The presence of $z$ in the denominator is natural as the down-jump
happens to one of the $z$ smaller triangles that touch the larger
triangle. What is the crucial invention of Wegner and Grossmann is
the {\it rate asymmetry} parameter $r$. It is this parameter that
makes the up-jumps (down-jumps) a predominant process if $r>1 (<1)$.

The {\it rate scaling} parameter $s$ is natural on every hierarchical
structure of the type Fig.~\ref{gasket} in order to compensate for
the increasing length of jumps as the level increases. The "normal"
case is when $s<1$. In this case the L\'evy flights cannot occur. On
the other hand, if $s>1$ and $r=1$, {\it all} the  moments $\langle
x^{m}\rangle$ ($m>0$) of random walk displacement are divergent at
all times. This is a somewhat pathological model.

The presence of the rate asymmetry parameter allows one to reach
new regime {\myred (regime ``C'' in Ref.[\onlinecite{Wegner-long}])} 
at $r< 1$ where $s>1$ but $(r\,s)<1$. It is exactly the regime
where the density correlation function behaves similarly to that in
the quantum case of random critical long-range Hamiltonians of
Eq.~(\ref{long-range}). Note that the case $r<1$ and $s=1$ would
result in trapping of a random walker on a fractal with no
restriction to the level number from below. Indeed, in this case the
bias towards down-jumps would drive the walker to lower and lower
levels of hierarchy without a considerable displacement in the
space. It is not the case in the WG model where the minimal level of
hierarchy does exist, so after reaching it, the walker must go to the
higher level with certainty and thereby may escape to arbitrary
large distance. Thus, the {\it coarse-grained} WG model {\it never
leads to localization} but only to {\it sub-diffusion}:
\begin{equation}
\label{sub-diffusion} \langle x^{2}\rangle \propto t^{\Theta}\equiv
t^{\frac{2}{d_{w}}},\;\;\;\;(\Theta<1).
\end{equation}
In Eq.~(\ref{sub-diffusion}) we introduced the {\it walk dimension}
$d_{w}$ which is essentially the {\it dynamical exponent} that
determines the relative scaling of space and time. In the WG model
with
\begin{equation}
\label{param} s>1,\;\;\;\;\; s r <1,
\end{equation}
it is given by \cite{Wegner-long}:
\begin{equation}
\label{walk-dim} d_{w}=\frac{\ln(s r)^{-1}}{\ln\mu}
\end{equation}
and leads to a sub-diffusion $\Theta<1$ if $(s r)^{-1}> \mu^{2}$.

The definition Eq(\ref{sub-diffusion}) is similar to the celebrated critical subdiffusion 
$ \, \langle x \rangle^2 \propto t^{2/d} \, $ which is well-known from the scaling theory of 
Anderson localization  \cite{GoF}. This suggests that $ \, d_w \, $ in (\ref{sub-diffusion}) 
has the same meaning as the dimensionality of space in the quantum critical models.
%
%
%
%
Now we show that the WG random walk with the
parameters $s$ and $r$ obeying Eq.~(\ref{param}) reproduces all the
regimes shown in Fig.~\ref{Levy}.

First of all, we note that, according to the Table~3 of
Ref.~[\onlinecite{Wegner-long}], the moments $\langle |x|^{m}\rangle$  
{\myred
are divergent  if $ \, \mu^m (r s) > 1 $ which requires $ \, {m>d_{w}} \, $
at $ \, r s < 1 < s $; here $ \, d_w \, $ is given by Eq.~(\ref{walk-dim})}
and $x$ is the
$x$-coordinate of the displacement vector $\bf R=\{x,
x_{1},x_{2},...x_{d-1} \}$ in the $d$-dimensional space in which the
fractal is embedded. This suggests that the probability to find a
walker at in an interval $(x,x+dx)$ at a time $t$ is proportional to
$|x|^{-(d_{w}+1)}\,dx$.

Switching to the radial-angle variables we find that the probability
to find the walker at a distance interval $\{ R,R+dR\}$ within the
solid angle interval $d\Omega$ is:
\begin{eqnarray}
\label{large-dist}P(R,t)&\propto&  R^{-d_{w}-1}\,R^{-d+1}\,R^{d-1} d
R d\Omega\\ \nonumber  &\propto & R^{-d}\,R^{-d_{w}}\,d^{d}{\bf
R},\;\;\;\;\;R\rightarrow\infty.
\end{eqnarray}
In order to find the coefficient of proportionality in
Eq.(\ref{large-dist}) we use Eq.(4.30) of Ref.[\onlinecite{Wegner-long}].
Then we obtain
\begin{equation}
\label{weg-P} R^{d}\,P(R,t)\equiv S(R,t)\approx
q^{(l)}_{l}(t)=1-e^{-\kappa\,(r s)^{l}\,t} \, .
\end{equation}
Here $\kappa=(1-r)(1-r s)$ and
$l\equiv l(R)=\ln R/\ln \mu$ is the displacement in the
ultra-metric space of levels expressed in terms of the radial
displacement in the real space. At large enough $R$ and $s r<1$ the
exponent can be expanded and we obtain:
\begin{equation}
\label{Levy-Weg} R^{-d}\,P(R,t)\equiv
S(R^{d_{w}}/t)=\kappa\,t\,R^{-\frac{\ln (r
s)^{-1}}{\ln\mu}}=\kappa\,t\,R^{-d_{w}}.
\end{equation}
We see that, in full agreement with the meaning of $d_{w}$ as an
exponent of dynamical scaling, $R^{-d_{w}}$ enters the function
$S(R,t)$ in the scale-invariant combination $R^{d_{w}}/t$. Thus
 Eq.(\ref{Levy-Weg}){\myred,  which describes} the Levy flights for the
classical WG random walk, is isomorphic to Eq.(\ref{PL-tail})  for
the density correlation function in the
quantum problem of random critical long-range Hamiltonians, provided
that:
\begin{equation}
\label{d-w-d} d_{w}\rightarrow d.
\end{equation}
Now consider small distances $R^{d_{w}}\ll t $. At such distances
Eq.(\ref{Levy-Weg})
no longer applies. In order to find the density correlation function in
this regime we
apply the scaling $S(R,t)=S(R^{d_{w}}/t)$ and the coarse-graining. The
latter implies that
the return probability $P(t)$ is proportional to $S(1,t)=S(1/t)$. So, the
time-dependence of
 return probability allows (via scaling) to find the entire function:
 \begin{equation}
 \label{return}
S(R,t)=S(R^{d_{w}}/t)\propto P(t/R^{d_{w}}).
 \end{equation}
The return probability in the WG random walk process{\myred, regime Eq.(\ref{param}),} 
has been found in Ref.[\onlinecite{ret-prob}]:
 \begin{equation}
 \label{ret-prob}
P(t)\propto t^{-\nu}, \;\;\;\;\;\nu=\frac{\ln
z}{\ln(rs)^{-1}}=\frac{d_{h}}{d_{w}},
 \end{equation}
where $d_{h}$ and $d_{w}$ are the Hausdorff and the walk dimensions given by
Eq.(\ref{Haus}) and Eq.(\ref{walk-dim}). Then using Eq.(\ref{large-dist}), and
Eqs.(\ref{return},\ref{ret-prob}), one finds for the density correlation function at
$R^{d_{w}}\ll t$:
\begin{equation}
\label{multi-Weg}
R^{-d}\,S(R,t)\propto R^{-d}\,\left(
\frac{R^{d_{w}}}{t}\right)^{\frac{d_{h}}{d_{w}}}.
\end{equation}
This is a classical counterpart of Eq.(\ref{MF-regime}). Quite naturally,
the correlation
fractal dimension $d_{2}$ is replaced by the Hausdorff dimension of the
classical fractal:
\begin{equation}
\label{dict-Haus}
d_{h}\rightarrow d_{2},
\end{equation}
and again the correspondence Eq.(\ref{d-w-d}) holds true.

We conclude this section by saying that there is a complete
quantum-to-classical analogy in
the density correlation function of the WG random walks and the long-range
critical random Hamiltonians.

\section{Conclusion}

The main results of this paper are the following:

{\bf (i)} We have identified two regions with qualitatively different behavior of the density correlation 
function of long-range critical Hamiltonians Eq.~(\ref{long-range}): the {\it multifractal region} where the
power-law behavior Eq.~(\ref{MF-regime}) predicted by Chalker \cite{Chalker} is valid and the {\it L\'evy
flights region} with the power-law behavior Eq.~(\ref{PL-tail}). Both types of behavior were
studied analytically within the virial expansion method and by a direct numerical
diagonalization of large matrices for the critical random-matrix model Eq.~(\ref{var}).

{\bf (ii)} It appears that for strong multifractality (
$b\ll 1$ in Eq.~(\ref{var})) there is
a complete analogy of the density correlation function in the quantum problem due to
{\it emergent fractality} of
random critical wave functions and in the classical random walks on hierarchical structures due to
their {\it geometrical fractality}. {\myred In both cases, one can find two independent critical exponents
in the density correlation function: Classical random walks on fractals can be described by
 the Hausdorff dimension $d_{h}$ and the walk dimension $d_{w}$ (or the spectral dimension 
$d_{s}=2d_{h}/d_{w}$).} 
%
%
{\myred Two corresponding critical exponents of the quantum problem are the correlation multifractal 
dimension $d_{2}$ and the dimensionality of space $d$.} 

%
%

{\bf (iii)} It is remarkable that only the latter determines the
power-law large-distance tail ({\myred quantum} Levy flights) in the scaling 
function {\myred $S(t \ll r^d) \propto t/r^{d} $, Eqs.(\ref{scaling},\ref{PL-tail}). The}
%
%
exponent $-d$ is independent of
$b$  (and hence on $d_{2}(b)$) at all values of the parameter $b$ in
Eq.(\ref{var}). {\myred By comparing Eq.(\ref{PL-tail}) and Eq.(\ref{Levy-Weg}),
we find that} the scaling variable $x=r^{d}/t$ (quantum critical
problem) or $x=r^{d_{w}}/t$ (classical random walks) enters in the
Levy flights power-law tail in both cases as $x^{-1}$ with the
universal exponent $-1$ which is independent of the correlation
dimension $d_{2}$ (quantum critical problem) or of the Hausdorff
dimension $d_{h}$ (classical random walks). We believe that this
universality has a deep physical origin.

\begin{figure}[ht]
\includegraphics[width=8.5cm, 
                             angle=0]{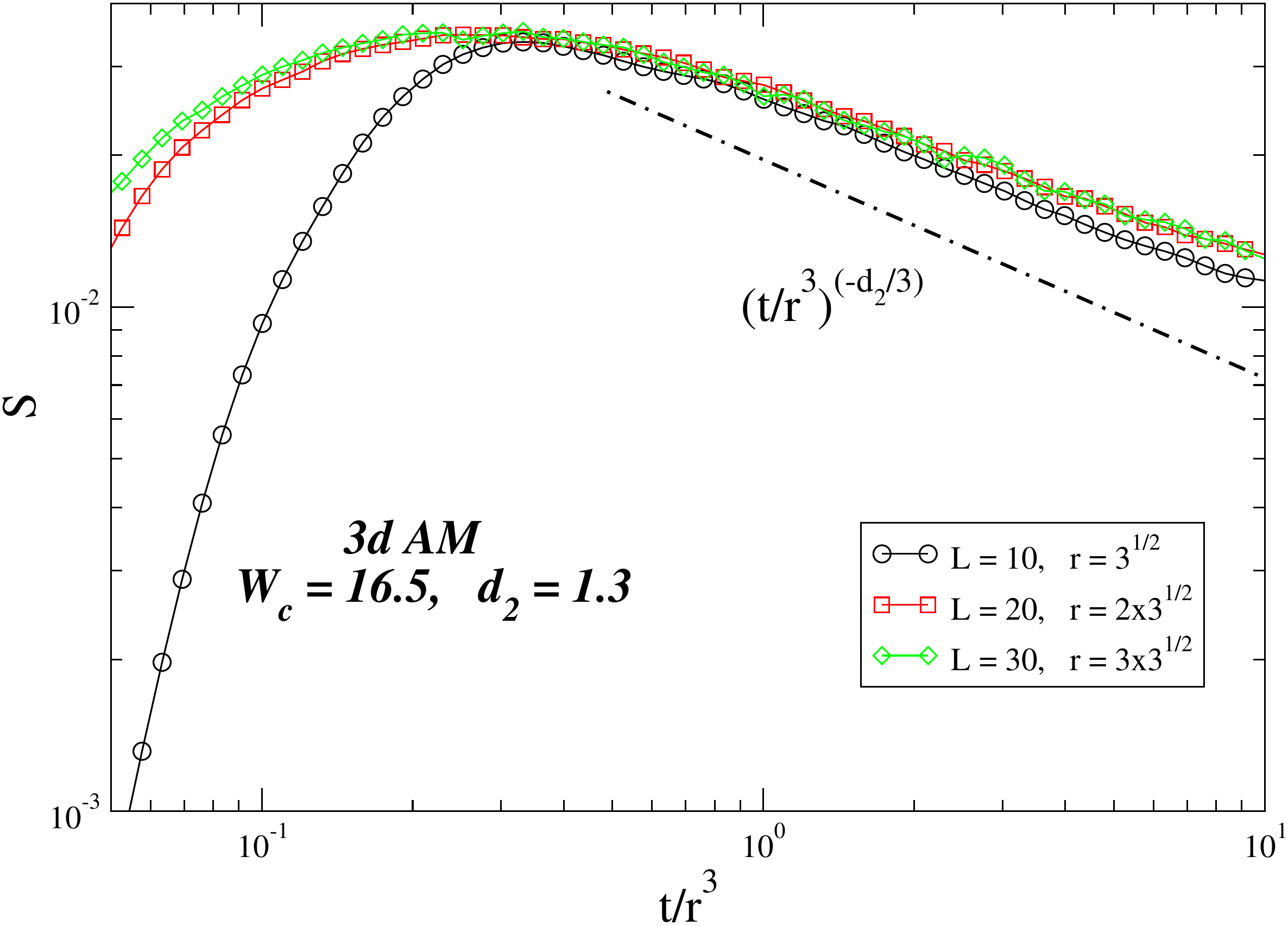}
\caption{(color online) 
%
%
{\myred Scaling function $ \, S = r^{3}\hat{\cal D}_{r,0}(t) \, $} vs. scaling variable
$t/r^{3}$ for the 3D Anderson localization model with the critical
disorder $W=16.5$ and the nearest-neighbor hopping integral $V=1$.
The plots for three different values of $r=3^{1/2}$, $r=2\times
3^{1/2}$ and $r=3\times 3^{1/2}$, and the corresponding three sample
sizes $L=10$, $L=20$ and $L=30$ collapse (at $t>t_{max}$) in the one
single scaling curve with the ``multifractal" power-law behavior
$(r^{3}/t)^{d_{2}/3}$, where $d_{2}\approx 1.3$ is the correlation
fractal dimension. The dotted-dashed line represents the
corresponding power-law. 
At $t<t_{max}$
there is a significant divergence of curves for different $r$. Given
that the ratio $r/L\approx 0.17$ is constant for all the curves,
this divergence can only be explained by a significant role played
by a ``microscopic length" $\ell_{0}$ (see Ref.[\onlinecite{KravCue}]) 
and the corresponding dimensionless combination
$(L_{t}/\ell_{0})^{3}=tE_{0}\sim t V=t$. Note that at maximum of the
scaling curve $t=t_{max}=1.5$, $12$ and $40$ for the three values of
$r$ indicated above. For the same values of $r$, the smallest value
of the scaling variable shown on the plot corresponds to $t=0.25$,
$2.0$ and $6.8$. The ``macroscopic limit" $L_{t}/\ell_{0}\gg 1$
implies $t\gg 1$. Only in this limit one expects a collapse of the
curves. This condition is obviously violated for the curve with the
smallest $r=3^{1/2}$ at $t<t_{max}$.} \label{AM}
\end{figure}

The application of the above results to the critical point of the
Anderson localization transition is a subtle issue due to the
absence of an exact solution to the Anderson localization problem
and due to the limited sizes of 3D systems amendable to direct
numerical diagonalization. Yet, it is rather well established that
the multifractal behavior of the density correlation function,
Eq.~(\ref{MF-regime}), is present at the Anderson transition point in
3D systems.
In Fig.~\ref{AM} we present the results of a direct numerical
diagonalization of the 3D Anderson model at critical disorder
$W=16.5$ and the nearest-neighbor hopping $V=1$. The density
correlation function $\hat{\cal D}_{r,0}(t)$ is computed for three
different values of $r$ and the corresponding three different system
sizes such that $r/L\approx 0.17={\rm const}$. Thus geometrically
the systems are ``macroscopically similar" and the finite-size
corrections of the type $r/L$, though appreciable, are the same for
all the three cases. One can see that for $t>t_{max}$ 
($ \, t_{max} \, $ corresponds to the maximum of the scaling curve)
all the three curves for $r^{3}\,\hat{\cal D}_{r,0}(t)$ collapse to the one single
scaling curve $S(r^{3}/t)\sim (r^{3}/t)^{d_{2}}$ with rather good
accuracy. In this region they exhibit the power-law behavior which
is represented by Eq.~(\ref{MF-regime}), albeit with the power
slightly modified by the finite-size $r/L$ effects.

The situation with the tail at $r^{3}\gg t$ is much worse. The
curves for different $r$ diverge significantly as the scaling
variable $t/r^{3}$ gets smaller. There is only one reason for such a
behavior: it is the finite-size effects of the type
$\ell_{0}/L_{t}$, where $\ell_{0}\sim (\rho_{0}V)^{-1/3} $
is the microscopic length which has been introduced in 
Ref.[\onlinecite{KravCue}], and $L_{t}^{3}=t/\rho_{0}$. 
Here $\rho_{0}\sim (a^{3}\,V\,W)^{-1}$ is
the density of states, {\myred thus $\ell_{0}\sim a\,W^{1/3}$
is of the order of the lattice constant $a$.}
The parameter that controls the smallness of
such finite-size effects is $(\ell_{0}/L_{t})^{3}\sim
(tV)^{-1}=1/t$. Obviously, at small $t$ this parameter is no longer
small and one cannot expect a scaling behavior in the region $t\ll
1$. On the other hand, $t_{max}\approx 0.3\,r^{3}$. So, it is only at
large $r$ that one may observe a scaling behavior in a sufficiently
wide interval $1\ll t<t_{max}$. We conclude that in order to meet
the requirement $1\ll t<t_{max}$ and simultaneously to exclude the
finite-size $r/L$ effects one should study {\it really} large 3D
systems with $L\sim 100$ which is numerically a very hard problem.

%
%

In this situation we can only guess what is the form of the scaling
function $S(r^{3}/t)$  at $t< t_{max}$ at the Anderson transition
point of 3D systems. The most natural assumption is that it is
exponentially small, and thus the Levy flights are absent, due to
the short-range nature of the Anderson Hamiltonian. A
counter-scenario in which the Levy flights may be present, is
offered \cite{AKL} by the observation of the "explosion of
high-gradient operators" in the non-linear supersymmetric
sigma-model in $(2+\epsilon)$ dimensions. Unfortunately, this
long-standing controversy still remains unresolved.

\acknowledgments
O.~Ye. acknowledges support from the DFG through grant SFB TR-12, the Arnold 
Sommerfeld Center for Theoretical Physics in Munich, and the Nanosystems Initiative 
Munich Cluster of Excellence and hospitality of the Abdus Salam ICTP where a part
of this project has been done. V.~K. and E.~C. acknowledge support  from the FEDER 
and the Spanish DGI through project No. FIS2010-16430.

\end{document}